\documentstyle[12pt]{article}

\begin{document}
\pagestyle{empty}
\begin{titlepage}

\vspace*{-8mm}
\noindent 
\makebox[9.9cm][l]{BUHEP-95-30} October 26, 1995\\
\makebox[9.9cm][l]{To appear in Phys.~Rev.~{\bf D}}
Revised: December 20, 1995\\
\makebox[9.9cm][l]{hep-ph/9510424}

\vspace{2.cm}
\begin{center}
 {\LARGE {\bf Gluino mass from dynamical \\ [2mm] supersymmetry
    breaking}}\\
\vspace{42pt}
{\large {\rm Bogdan A. Dobrescu}}\footnote{e-mail address:
dobrescu@budoe.bu.edu}
\vspace*{0.6cm}

{\it Department of Physics, Boston University \\
590 Commonwealth Avenue, Boston, MA 02215, USA}

\vskip 3.4cm
\end{center}
\baselineskip=18pt

\begin{abstract}

{\normalsize
We present a new mechanism for gluino mass generation in models of
dynamical supersymmetry breaking. The mechanism requires two colored
chiral superfields which feel a nonabelian gauge interaction such that
a fermion condensate is formed at a scale of order 1 TeV.
Renormalizable hidden sector models, which typically yield
unacceptably light gauginos, become viable if a gauge singlet is
coupled to these chiral superfields. Moreover, the interactions of the
gauge singlet with the Higgs superfields substitute the $\mu$-term.
Visible sector models can also incorporate this mechanism; however,
the models of dynamical supersymmetry breaking previously analyzed
cannot be fitted in a simple visible sector model because they lead to
vacuum expectation values for charged scalars.
}

\end{abstract}

\vfill
\end{titlepage}

\baselineskip=18pt
\pagestyle{plain}
\setcounter{page}{1}

\section{Introduction}
\label{sec:intro}
\setcounter{equation}{0}

Supersymmetry (SUSY)
ensures the stability of the electroweak scale against radiative
corrections (i.e.~it solves the naturalness problem) without
explaining the small ratio of the electroweak scale to the Planck
scale (the hierarchy problem).
However, the hierarchy problem is solved
if supersymmetry is broken dynamically, i.e.~the vacuum is
non-supersymmetric due to non-perturbative effects of some
non-abelian gauge interaction, called supercolor.

There are known three classes of candidates for the supercolor sector
(we include here only renormalizable models of calculable dynamical
SUSY breaking): the $SU(3)\times SU(2)$
model \cite{ads1} and its generalizations \cite{dnns},
the $SU(2n+1)$ models \cite{ads1,ads2}, and the $SU(2n)\times
U(1)$ models \cite{dnns, erich}. All these models
have a spontaneously broken $R$-symmetry and, apparently, a massless
$R$-axion.
However, Bagger, Poppitz and Randall \cite{axion}
showed that the cancellation of the cosmological constant implies
the explicit breaking of the $R$-symmetry such that the $R$-axion
is massive.

The outstanding phenomenological issue in this framework is to
find a viable way of transmitting SUSY breaking from the supercolor
sector to the minimal supersymmetric standard model (MSSM).
In this approach it is difficult to produce a sufficiently
large gluino mass, i.e.~of the order of the weak scale (the
possibility that light gluinos, with mass of order 1 GeV, are not
experimentally ruled out is controversial \cite{data}).
The same problem arises in models with a continuos $R$-symmetry
\cite{rsym}.

In the case of hidden sector models, SUSY breaking is transmitted
by supergravity from the hidden sector (which includes supercolor
interactions) to the visible sector (which includes the MSSM).
Nonrenormalizable interactions suppressed by powers of the Planck
scale, $M_P$, give masses to the scalars and gauginos of the  visible
sector. The scalar masses in the visible sector are of  order
$\Lambda_{SC}^2/M_P$, where $\Lambda_{SC}$
is the scale where the supercolor
interaction becomes strong. Scalar masses of the order of the weak
scale correspond to $\Lambda_{SC}  \sim 10^{11}$ GeV.
The gluino and photino masses, however, are produced by
higher-dimensional operators, so they are suppressed by at least two
powers of $M_P$ and are unacceptably small \cite{ads1,gluino}.

The situation is improved if there are light color-octet chiral fields
which can mix with the gluinos \cite{rsym}. However, such mixing is
associated with a large breaking of hypercharge unless
hypercharge is unified in a non-abelian group broken at an
intermediate scale \cite{gluino}.
Also, additional fields are required to enhance the photino mass.
Thus, this alternative is very complicated.

Another possibility is that the hidden sector contains a singlet
whose $F$-term acquires a VEV \cite{dsb}.
In this case, the gluino mass is
suppressed by only one power of $M_P$. However, the supercolor sector
has chiral content, so that the $F$-term for the singlet can only be
induced at one or two loops. As a result, the gluino mass is still
smaller than the weak scale by at least one order of magnitude.

In visible sector models, SUSY breaking is communicated to the MSSM
by gauge interactions.
Affleck, Dine and Seiberg \cite{ads1} constructed the simplest scheme
of this type by gauging a global symmetry of the supercolor sector and
identifying it with one of the gauge symmetries of the standard model.
However, they found that in order to give a sufficiently large mass to
the gluino the gauged symmetry should be $SU(3)_C$, such that color
becomes strong above the SUSY breaking scale.
Dine and Nelson \cite{dn} proposed a more sophisticated scheme which
allows a large gluino mass without major phenomenological problems.
This involves a new gauge interaction (the ``messenger group'') which
communicates SUSY breaking to some additional fields. These have
interactions with a gauge singlet which obtains a VEV.
Finally, some vector-like quarks and leptons have a non-supersymmetric
spectrum due to the coupling to the gauge singlet. Subsequent versions
of this approach were greatly
simplified, firstly \cite{dns} by taking advantage of the observation
that the $R$-axion is massive and secondly \cite{dnns} by avoiding
Fayet-Iliopoulos terms for the messenger $U(1)$.
The result is a highly predictive model with all the soft breaking
parameters of the MSSM determined in terms of only two unknown
parameters.
However, a large number of fields is still required and the
problem of electroweak symmetry breaking is not settled.

The above approaches show that it is possible to construct
realistic models of dynamical SUSY breaking but one has to introduce
very complicated structures for avoiding light gluinos. Hence, it
would be useful to find simpler methods for gluino mass generation.

In this paper we present a new mechanism for producing a gluino mass.
The idea is to introduce a new gauge interaction and to arrange that
a gluino mass arises dynamically when this interaction becomes
strong. This has similar features to the fermion mass generation in
technicolor models \cite{tc}. This mechanism for gluino mass
generation is quite general and can be useful in hidden or visible
sector models of dynamical SUSY breaking, or in models with
a continuous $R$-symmetry.

In section 2 we discuss the mechanism in general, without considering
the origin of SUSY breaking. Sections 3 and 4 specialize this
mechanism to hidden sector models and visible sector models,
respectively. A summary of the results and some final remarks
are given in section 5.

\section{Gluino Mass}
\label{sec:gluino}
\setcounter{equation}{0}

Consider the MSSM with massless gluinos at tree level.
The radiative gluino masses are small, of order 1 GeV \cite{oneloop}.
Let us introduce a new non-abelian gauge interaction (``$G$-color'')
which becomes strong at a scale $\Lambda_G$ larger than the weak
scale. For simplicity we take this to be
$SU(N_G)$. We also introduce two chiral superfields,
$\psi_L$ and $\psi^c$, with $SU(N_G)\times SU(3)_C\times SU(2)_W\times
U(1)_Y$ quantum numbers:
\begin{equation}
\psi_L: (N_G, 3, 1)_y  \; \; \; , \; \; \; \;
\psi^c: (\overline{N_G}, \overline{3}, 1)_{-y}~.
\end{equation}
The asymptotic freedom of QCD is preserved if $N_G = 2$;
$N_G = 3$ or 4 is also acceptable, since
the QCD coupling constant remains weak below the Planck scale
($\alpha_s(M_P) \begin{array}{c}\,\sim\vspace{-21pt}\\< \end{array}
0.25$ for $N_G = 4$).

The notations we use for a superfield and for its
upper-spin component are the same
while the lower spin component is distinguished by a tilde.
The only exception will be that we use the standard notation for
Higgs fields (i.e. the tilde is on higgsinos).

We assume that the $G$-colored scalars have positive squared masses,
$M^2_{\tilde{\psi}_L}$ and $M^2_{\tilde{\psi}^c}$, 
and that there is a dynamically generated mixing
\begin{equation}
M^2_{LR} \tilde{\psi_L}\tilde{\psi^c} + {\rm h.c.}
\end{equation}
Such soft SUSY breaking terms arise in models of
dynamical SUSY breaking under certain
conditions which are discussed in the following two sections.

If the scalar masses are larger than $\Lambda_G$, the low energy
$G$-color theory consists of $G$-gluinos and $N_f = 3$ massless
flavors of $G$-fermions in the fundamental representation.
In the absence of scalar mixing, this theory has a non-anomalous
$U(1)_R$ symmetry \cite{ads3} in addition to the usual chiral symmetry
$SU(3)_L \times SU(3)_R \times U(1)_V$.
Under this $R$-symmetry, the scalars $\tilde{\psi_L}$ and
$\tilde{\psi^c}$ have charge $1 - N_G/N_f$ (in the normalization
were the Grassmann variable $\theta$ has charge $-1$),
so that their mass mixing violates $U(1)_R$. This is important in what
follows because the gauginos have $R$-charge $+1$ and, thus,
any gaugino mass violates $U(1)_R$ by $+2$ units.

The main effect of scalar mixing is given by a dimension-six operator
involving two $G$-gluinos, $\tilde{G}$,
which arises due to the scalar exchange diagram shown in fig. 1:
\begin{equation}
\frac{2 g_G^2}{M^2_{\tilde{\psi}}} 
(\tilde{G}^a\psi_L)T^a T^b(\psi^c\tilde{G}^b)~,
\label{eq:op}
\end{equation}
where $T^a$ are the generators of the fundamental representation,
$g_G$ is the $G$-color gauge coupling and
\begin{equation}
M^2_{\tilde{\psi}} = \frac{M^2_{\tilde{\psi}_L} 
M^2_{\tilde{\psi}^c} - M^4_{LR}}{M^2_{LR}} ~.
\label{mpsi}
\end{equation}

\begin{picture}(220,130)(-35,-10)
\thicklines
\put(40,7.5){\line(4,3){50}}
\thinlines
\put(65,26.5){\vector(4,3){0}} \put(68,14){$\tilde{G}$}
\put(40,82.5){\line(4,-3){50}}
\put(65,63.5){\vector(4,-3){0}} \put(65,70){$\psi_L$}
\put(90,45){\line(1,0){15}}
\put(115,45){\line(1,0){15}}
\put(125,45){\vector(1,0){0}} \put(114,55){$\tilde{\psi_L}$}
\put(140,45){\line(1,0){20}}
\put(170,45){\line(1,0){15}}
\put(175,45){\vector(-1,0){0}} \put(177,55){$\tilde{\psi^c}$}
\put(195,45){\line(1,0){15}}
\thicklines
\put(144,39){\line(1,1){12}} \put(144,51){\line(1,-1){12}}
\put(210,45){\line(4,-3){50}}
\thinlines
\put(235,26.5){\vector(-4,3){0}} \put(222,14){$\tilde{G}$}
\put(210,45){\line(4,3){50}}
\put(235,63.5){\vector(-4,-3){0}} \put(225,68){$\psi^c$}
\end{picture}

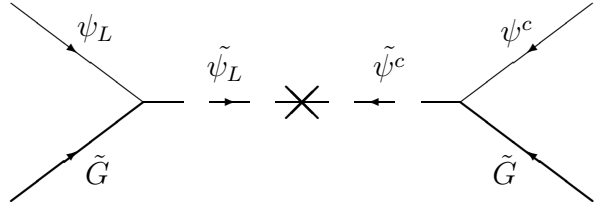
\hspace{2ex} Fig. 1. {\small \ Scalar exchange diagram. The cross
  represents scalar mixing.}

\vspace{2ex}

When the $G$-interaction becomes strong, the fermions condense.
It is possible that a $\langle\tilde{G}\tilde{G}\rangle$ or
$\langle\psi_L\tilde{G}\psi^c\tilde{G}\rangle$ condensate forms, as
suggested in ref. \cite{exotic}.
However, it is more likely that
a $\langle\psi_L\psi^c\rangle$ condensate forms, as in non-SUSY QCD.
This is due to the operator (\ref{eq:op}): if
the fermion bilinear has a small VEV, a $G$-gluino mass is generated
and the theory tends to become similar to non-SUSY QCD. Therefore, the
condensate could
increase up to the point where the $G$-gluino mass is
larger than $\Lambda_G$. If this is the case, the $G$-gluino can be
integrated out and the theory becomes a scaled up version of
ordinary QCD with three flavors, so that chiral symmetry is broken
by
\begin{equation}
\langle\psi_L\psi^c\rangle \approx \Lambda_G^3~.
\label{eq:cond}
\end{equation}

To check the consistency of this scenario, we compute the $G$-gluino
mass by Fierz transforming the operator (\ref{eq:op})
and using eq. (\ref{eq:cond}):
\begin{equation}
M_{\tilde{G}} \approx \frac{2 \pi \alpha_G \Lambda_G^3}
{N_G M^2_{\tilde{\psi}}}~,
\label{eq:mG}
\end{equation}
where $\alpha_G = g_G^2/(4\pi) \sim {\cal O} (1)$.
Note that the $R$-charge of the $G$-fermions is $-N_G/N_f$
such that the $R$-charge of the $G$-gluino mass term,
\begin{equation}
R( \tilde{G}\tilde{G} ) =
R( \langle\bar{\psi_L}\bar{\psi^c}\rangle )
+ R( \tilde{\psi_L}\tilde{\psi^c} )~,
\end{equation}
is indeed $+2$. Eq. (\ref{eq:mG}) shows that it is not possible
without fine-tuning to have both the $G$-gluinos and the $G$-scalars
much heavier than the $G$-color scale. We consider the case were
$M_{\tilde{\psi}_L}, M_{\tilde{\psi}^c}, 
M_{LR} \sim {\cal O} (\Lambda_G)$ so that
$M_{\tilde{G}} \sim {\cal O} (\Lambda_G)$.
In this case, the low energy $G$-color dynamics might be influenced by
the $G$-scalars and $G$-gluinos. However, this effect is
probably not large enough to change the structure of the vacuum and
we will neglect it in the rough estimate of the gluino mass.

The reversed argument, that a $G$-gluino condensation in connection
to the operator (\ref{eq:op}) could lead to a large mass for the
$G$-fermions, does not apply because the chirality-flip scalar mixing
requires the $G$-fermion condensate. Still, it is possible that both
the $G$-fermion condensate and the $G$-gluino condensate appear.
This would induce a scalar mixing of order $\alpha_G \Lambda_G^2$,
as shown in fig.~2.


\begin{picture}(200,115)(-80,50)

  \put(100,100){\oval(40,40)[l]}
  \put(114,100){\oval(40,40)[r]}
  \put(107,120){\circle*{14}}
  \put(107,80){\circle*{14}}
  \put(80,100){\line(-1,0){10}} \put(65,100){\line(-1,0){10}}
  \put(50,100){\line(-1,0){10}} \put(35,100){\line(-1,0){10}}
  \put(134,100){\line(1,0){10}} \put(149,100){\line(1,0){10}}
  \put(164,100){\line(1,0){10}} \put(179,100){\line(1,0){10}}

  \put(72,120){$\psi_L$}
  \put(130,120){$\psi^c$}
  \put(75,70){$\tilde{G}$}
  \put(127,70){$\tilde{G}$}
  \put(35,110){$\tilde{\psi_L}$}
  \put(170,110){$\tilde{\psi^c}$}

\end{picture}

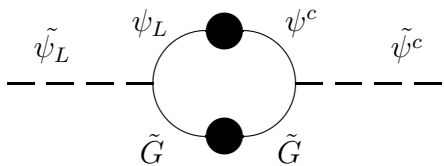
\hspace{2ex} Fig.~2. {\small Scalar mixing due to the
 $\langle\tilde{G}\tilde{G}\rangle$ {\it and}
$\langle\psi_L\psi^c\rangle$ condensates (represented by $\bullet$).}

\vspace{2ex}

\noindent
Whether or not this happens, the discussions that follow will not be
affected since they rely only on the assumption that the $G$-fermions
condense.

Now we have the tools for producing a gluino mass. The scalar exchange
diagram of fig.~1 with the external $G$-gluinos replaced by
ordinary gluinos yields an effective four-fermion interaction,
which gives a gluino mass when the $G$-fermions condense:
\begin{equation}
M_{\tilde{g}} \approx \frac{2 \pi \alpha_s \Lambda_G^3}
{3 M^2_{\tilde{\psi}}}~,
\label{gluino}
\end{equation}
where $\alpha_s \approx 0.1$ is the strong coupling constant
at the scale $\Lambda_G$. A gluino mass of few hundred GeV can be
easily obtained with $\Lambda_G \sim {\cal O} (1~{\rm TeV})$.
Note that the experimental lower bound for the gluino mass
is model dependent and in the range $100 - 220$ GeV \cite{data}.

Similarly, the diagram of fig.~1, with photinos instead of
$G$-gluinos, gives a Majorana mass for the photino:
\begin{equation}
M_{\tilde{\gamma}} = \frac{\pi \alpha y^2}{\cos^2\theta_W}
\frac{\Lambda_G^3}{M^2_{\tilde{\psi}}}~.
\end{equation}
where $\alpha$ is the fine structure constant and $\theta_W$
is the weak angle.
The condition of having the Landau pole of $U(1)_Y$ above
$M_P$ imposes a bound on the hypercharge of $\psi_L$,
\begin{equation}
|y| \begin{array}{c}\,\sim\vspace{-21pt}\\< \end{array} 1
\label{eq:y}
\end{equation}
(we use the convention
$Y/(T_3 - Q) \equiv 2$, where $Y$ is the hypercharge,
$Q$ is the electric charge and $T_3$ is an $SU(2)_W$ generator).
Thus, there is an upper bound on the
ratio of the Majorana masses of the photino and gluino:
\begin{equation}
\frac{M_{\tilde{\gamma}}}{M_{\tilde{g}}} = \frac{3 \alpha y^2}
{2 \alpha_s \cos^2\theta_W} 
\begin{array}{c}\,\sim\vspace{-21pt}\\< \end{array} 0.15 ~.
\label{photino}
\end{equation}

\section{Hidden Sector Models}
\label{sec:sugra}
\setcounter{equation}{0}

If supersymmetry is broken dynamically  in a hidden sector
by supercolor interactions at a scale $\Lambda_{SC} \sim 10^{11}$ GeV,
all the scalars get masses of order 1 TeV at this scale,
while the gaugino masses are suppressed at least by an
additional factor of $\Lambda_{SC}/M_P$.

In order to use the mechanism for gluino mass generation presented
in the previous section, there is need for scalar mixing. Such a
mixing could be produced by supergravitational interactions if the
Kahler potential were non-minimal.
Nevertheless, this situation is not likely to occur since the
constraints from flavor-changing neutral currents suggest that
the part of the Kahler potential responsible for squark and slepton
masses is minimal.

In the rest of this section we present a more natural source of
$G$-scalar mixing.

\subsection{$G$-scalar mixing}

We consider a gauge singlet, S, in the visible sector with
interactions described by the following superpotential:
\begin{equation}
W = \lambda_1 S \psi_L \psi^c + \frac{\lambda_2}{3} S^3 ~.
\end{equation}
When the $G$-fermions condense, the Yukawa interaction of the
$\tilde{S}$ scalar with the $G$-fermions gives rise to  a tadpole
term for the gauge singlet scalar.
The potential for $\tilde{S}$ is given by:
\begin{equation}
V_S = \lambda_2^2 |\tilde{S}|^4 +
M_{\tilde{S}}^2 |\tilde{S}|^2
+ \lambda_1 \Lambda_G^3 (\tilde{S} + \tilde{S}^{\dagger})
\end{equation}
The scalar $\tilde{S}$ has positive mass squared at the scale
$\Lambda_{SC}$, $M^2_{\tilde{S}} \sim {\cal O} \left((1
~{\rm TeV})^2\right)$,
but one loop corrections can drive $M^2_{\tilde{S}}$ negative at the
scale $\Lambda_G$ if $\lambda_1$ is large enough.
The minimum of the potential is at
\begin{equation}
\langle\tilde{S}\rangle =
- \left(\frac{\lambda_1}{4\lambda_2^2}\right)^{1/3}
\Lambda_G f(a) ~,
\label{svev}
\end{equation}
where
\begin{equation}
a \equiv \frac{2^{1/3}}{3 (\lambda_1\lambda_2)^{2/3}}
\, \frac{M_{\tilde{S}}^2}{\Lambda_G^2}
\label{alfa}
\end{equation}
and
\begin{equation}
f(a) = \left\{ \begin{array}{rcl}
\left[(1 + a^3)^{1/2} + 1 \right]^{1/3}
  - a \left[(1 + a^3)^{1/2} + 1 \right]^{-1/3}
\; & , & a \geq - 1 \\ [3mm]
2 |a|^{1/2} \cos \left( \frac{1}{3} \arccos |a|^{-3/2}
\right) \; \hspace{5em}  & , & a < - 1 \\ \end{array}\right.
\label{falfa}
\end{equation}
For Yukawa coupling constants $\lambda_1, \lambda_2 \sim {\cal O}
(1)$, eqs.~(\ref{svev})-(\ref{falfa}) give a negative VEV for
$\tilde{S}$ of order $\Lambda_G$, which is quite insensitive
to the sign or value of $M^2_{\tilde{S}}$ as long as the ratio
$|M^2_{\tilde{S}}|/\Lambda_G^2$ is not very large.
Due to this VEV, the Yukawa interaction of the $\tilde{S}$
scalar with $S$ fermions induces a Majorana mass for $S$:
\begin{equation}
m_S = \lambda_2 |\langle\tilde{S}\rangle| \sim {\cal O} (\Lambda_G) ~.
\end{equation}

The $S$ fermion exchange diagram shown in fig.~3 yields a
dimension-five operator,
\begin{equation}
\frac{\lambda_1^2}{m_S} \tilde{\psi_L}\tilde{\psi^c}\psi_L\psi^c ~,
\end{equation}
which gives a scalar mixing when the fermions condense:
\begin{equation}
M_{LR}^2 \approx  \lambda_1^2 \frac{\Lambda_G^3}{m_S}~.
\end{equation}

\begin{picture}(400,130)(-80,-30)
\put(0,0){\line(4,3){12}}
\put(18,13.5){\line(4,3){12}}
\put(24,18){\vector(4,3){0}} \put(34,10){$\tilde{\psi_L}$}
\put(36,27){\line(4,3){12}}
\put(0,72){\line(4,-3){48}}
\put(24,54){\vector(4,-3){0}} \put(34,60){$\psi^c$}
\put(48,36){\line(1,0){100}}
\put(98,36){\vector(-1,0){0}} \put(93,45){$S$}
\put(148,36){\line(4,3){48}}
\put(172,54){\vector(-4,-3){0}} \put(160,60){$\psi_L$}
\put(148,36){\line(4,-3){12}}
\put(166,22.5){\line(4,-3){12}}
\put(172,18){\vector(-4,3){0}} \put(155,10){$\tilde{\psi^c}$}
\put(184,9){\line(4,-3){12}}
\end{picture}

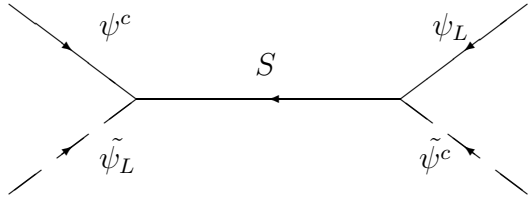
\hspace{2ex} Fig. 3. {\small \ Fermion exchange diagram responsible
      for scalar mixing.}

\vspace{2ex}

\noindent
Therefore, a Majorana gluino mass is generated as discussed in section
2 (see eqs.~(\ref{mpsi}) and (\ref{gluino})).
Note that this is the gluino mass at a scale of order 1 TeV.
At a scale of 200 GeV,
the renormalization group evolution gives the gluino mass
larger by a factor of about 1.2.
If all the mass parameters are of order 1 TeV and the Yukawa
couplings are of order one, then the gluino mass is of order
200 GeV, which is close to the experimental lower bound.

By contrast, if the Kahler potential is minimal, we expect
the squark and slepton masses to be in the $0.5 - 1$ TeV range.
Furthermore, since the gaugino masses are produced at the
low scale $\Lambda_G$, the renormalization group equations
are different than in the MSSM where it is
assumed that the gluino and photino masses are produced at a
high scale $\sim \Lambda_{SC}$.
In the MSSM the squarks are typically heavier than the sleptons
because there is a positive contribution at one-loop proportional
to the gluino mass squared. In the present context, this contribution
appears only below $\Lambda_G$ and is negligible. Therefore, the
squark and slepton masses will be almost equal. Only the
top scalars will be lighter due to negative one-loop contributions
proportional to the square of the large top Yukawa coupling.

\subsection{Linking the $G$-color scale with the $G$-scalar
masses.}

There is an artificial\footnote{I thank Lisa Randall for
emphasizing this point.} ingredient in the above discussion:
the scale associated with the $G$-scalar masses and the scale
of the $G$-color interaction have the same order of magnitude.
Both these scales arise naturally due to nonperturbative dynamics, but
the fact that they are close to each other requires an explanation.
Note that a similar problem \cite{link}
appears in supersymmetric technicolor models \cite{susytc}.

Since $M_{\tilde{\psi}_L}$ and $M_{\tilde{\psi}^c}$ 
have the same origin as the squark and slepton
masses, we do not expect them to be larger than about 1 TeV.
If $\Lambda_G$ were significantly smaller than $M_{\tilde{\psi}_L}$ 
(we assume $M_{\tilde{\psi}_L} = M_{\tilde{\psi}^c}$), then the gluino
mass would be too small, as can be seen from eq.~(\ref{gluino}).
If $\Lambda_G$ were much larger than $M_{\tilde{\psi}_L}$, 
$G$-color would probably be
spontaneously broken by a $G$-scalar VEV when $N_G \geq 3$
or the chiral symmetry would be unbroken when $N_G = 2$ \cite{exotic}.

The equality of these two scales can be seen as a fine-tuning of the
$G$-color coupling constant. Note that this is not a fine-tuning in
the technical sense (it is natural in the sense of 't Hooft \cite{nat}
to adjust a gauge coupling), but rather in the colloquial sense,
i.e. the range allowed for the $G$-color coupling constant is small.
The renormalization group evolution of the $G$-color coupling constant
between $\Lambda_G$, where $\alpha_G(\Lambda_G) \approx 1$, and
$M_{\tilde{\psi}_L}$ 
gives the range allowed for $\alpha_G(M_{\tilde{\psi}_L})$.
The one-loop coefficient of the $\beta$-function is
$b_0 = - 3 N_G + 2$ since the $G$-scalars decouple below
$M_{\tilde{\psi}_L}$.
For example, if we require $1 \leq M_{\tilde{\psi}_L}/\Lambda_G \leq
2$, then $0.56 
\begin{array}{c}\,\sim\vspace{-21pt}\\< \end{array} 
\alpha_G(M_{\tilde{\psi}_L}) 
\begin{array}{c}\,\sim\vspace{-21pt}\\< \end{array} 1$ for $N_G = 3$.
Although this does not seem to be a severe constraint,
the range of $\alpha_G$ shrinks at higher scales.
Using $b_0 = - 3 N_G + 3$ with $N_G = 3$ between $M_{\tilde{\psi}_L}$ 
and the
supercolor scale gives a 4\% fine-tuning in $\alpha_G(\Lambda_{SC})$
(we define the
amount of fine-tuning by $\alpha_G(\Lambda_{SC})_{\rm max}/
\alpha_G(\Lambda_{SC})_{\rm min} - 1$).

Such a fine-tuning might be explained if supercolor and $G$-color
are unified at a high scale. This would provide a connection
between the supercolor coupling constant,
$\alpha_{SC}(\Lambda_{SC}) \approx 1$, and $\alpha_G(\Lambda_{SC})$.

A more convenient approach is to avoid excessive fine-tuning
by slowing down the running of $\alpha_G$ above $M_{\tilde{\psi}_L}$. 
This requires additional $G$-colored chiral superfields.
As an example, consider the case $N_G = 3$. We introduce a new
chiral superfield in the adjoint representation of the $G$-color
group. Its scalar component receives a mass of order
$M_{\tilde{\psi}_L}$ from
supergravitational interactions while the fermion component condenses
when $G$-color becomes strong
(we assume that this does not prevent the formation of the
$\langle\psi_L\psi^c\rangle$ condensate).
The $\beta$-function above $M_{\tilde{\psi}_L}$ is given by
\cite{twoloop}:
\begin{equation}
\beta (g_G) \approx \frac{g_G^3}{16 \pi^2} \left( -3 + 2.7 \alpha_G
+ {\cal O} (\alpha_G^2) \right)~.
\end{equation}
We see that the asymptotic freedom is preserved while the rate of
running for $\alpha_G \sim {\cal O} (1)$ is very small. However,
previous studies of gauge theories with a slowly-running coupling
constant in the context of walking technicolor \cite{walk} show that
the convergence of the $\beta$-function is not reliable in this case.
Therefore, it is not very useful to compute the amount of fine-tuning
at the scale $\Lambda_{SC}$ using the two-loop $\beta$-function.
We expect, though, that the range of $\alpha_G$
will not shrink dramatically above $M_{\tilde{\psi}_L}$, while the
range of $\alpha_G(M_{\tilde{\psi}_L})$ 
will not be reduced significantly
with the inclusion of the additional superfield.

In conclusion, if the $G$-color coupling constant evolves
slowly, then the two scales, $M_{\tilde{\psi}_L}$ 
and $\Lambda_G$, are linked
\cite{link}: when the $G$-scalars decouple, $\alpha_G$ starts to
increase faster, triggering chiral symmetry breaking.

\subsection{Electroweak symmetry breaking}

Since we try to construct a viable way of transmitting dynamical SUSY
breaking, we have to specify the mechanism for electroweak symmetry
breaking.
At the scale $\Lambda_{SC}$ supergravitational interactions give
positive squared masses to the Higgs scalars, $H_u$ and $H_d$,
of order $M_{\tilde{\psi}_L}$.
Due to the large Yukawa coupling of the top quark, radiative
corrections drive $M_{H_u}^2$ negative at the scale
$M_{\tilde{\psi}_L}$ and
the electroweak symmetry breaks \cite{ewsb}.
Though, additional features are necessary: $H_d$ should also have a
VEV and the Higgsinos should be massive.
In the MSSM, these requirements are satisfied by a $\mu H_u H_d$
term in the superpotential.
In the present context it is not reasonable to introduce a
$\mu$-term since the main goal of dynamical SUSY breaking is
to explain the existence of any mass parameter other than $M_P$.

However, we can introduce a term
\begin{equation}
W^\prime = \lambda_3 S H_u H_d
\end{equation}
in the superpotential.
The potential for $\tilde{S}$ is modified in this case and we have
to minimize it simultaneously with respect to $\tilde{S}$,
$\tilde{H_u}$ and $\tilde{H_d}$. However, for simplicity
we consider the case were the Yukawa coupling constant is small,
$\lambda_3 \sim 0.2 - 0.3$, so that eqs.~(\ref{svev})-(\ref{falfa})
remain valid.
The VEV of $\tilde{S}$ then gives a higgsino
mass term,
\begin{equation}
- \lambda_3 \langle\tilde{S}\rangle \tilde{H_u} \tilde{H_d}~,
\end{equation}
and the VEV of the $F_S$ auxiliary field gives a scalar mixing
$B$-term,
\begin{equation}
- \lambda_2\lambda_3 \langle\tilde{S}\rangle^2 H_u H_d
+ {\rm h.c.}
\end{equation}
If $\langle\tilde{S}\rangle \sim {\cal O} (1~{\rm TeV})$ 
and $\lambda_2 \sim
{\cal O} (1)$, then the higgsino mass  is in the range of few hundred
GeV,
while the mass coefficient of the $B$-term is larger by a factor of 2
or 3.
Therefore, the gauge singlet can perform the tasks of a $\mu$-term.

Since the Majorana wino  mass is zero at tree level, the lightest
chargino could have the mass above the experimental lower limit of
$\sim 45$ GeV only if $\tan \beta \equiv \frac{v_2}{v_1}$ is small
($\tan \beta \begin{array}{c}\,\sim\vspace{-21pt}\\< \end{array} 3$) 
\cite{higgs}. The lightest neutralino has a mass
of order $M_{\tilde{\gamma}}$,
which is $\sim 30$ GeV when $M_{\tilde{g}} \sim 200$ GeV
and $y \approx 1$ (see eq.~(\ref{photino})); this is also
close to the experimental lower limit of 20 GeV \cite{data}.
It is possible, however, to give Majorana masses of order
$\sim 100$ GeV to the winos and zinos if there are two
flavors of $G$-colored superfields which form $SU(2)_W$ doublets.
Note that $\alpha_G$ runs slowly if we introduce four (six)
new flavors when $N_G = 3$ ($N_G = 4$).
In this case, the masses of the lightest neutralino and chargino
will increase.

We have to address a problem associated with gauge singlets.
A quadratically divergent tadpole can appear at two loops
in supergravity theories which contain a singlet under any
gauge and global symmetry \cite{singlet}.
This tadpole leads to a large VEV for the singlet which
destabilizes the mass hierarchy
if the singlet has renormalizable couplings to the  visible
sector.
Thus, in the present context this problem is avoided only if
$S$ is charged under a global symmetry.
The term $S^3$ from the superpotential breaks any such continuous
global symmetry (excepting an $R$-symmetry under which $S$ has
charge $\frac{2}{3}$; this is broken by the $S\psi_L\psi^c$ term).
Still, the superpotential $W + W^\prime$ has a discrete $Z_3$
symmetry. Since $S$ has non-zero $Z_3$ charge, the Higgs
fields are also charged so that the quarks and leptons have to carry
$Z_3$ charge. This symmetry can allow the most general couplings of
the quarks and leptons to the Higgs fields,
but also it can be used as a horizontal symmetry which
restricts these couplings (this would make a
connection between fermion masses and supergravity).

$Z_3$ is dynamically broken by the $G$-fermion condensate
which implies that
domain walls were produced when the Universe had a
temperature of order $\Lambda_G$ \cite{wall}.
As emphasized in refs.~\cite{dnns,dn},
this may not be a phenomenological problem
since there are scenarios leading to a sufficiently fast
decay of the domain walls: for example,
the $Z_3$ can be a remnant of a
spontaneously broken continuous symmetry \cite{wall,decay}.

\section{Visible Sector Models}
\label{sec:visible}
\setcounter{equation}{0}

Visible sector models are appealing for several reasons:
since SUSY breaking is transmitted by gauge interactions,
squark and slepton degeneracy arises naturally;
the coefficients of the soft SUSY breaking terms appearing in the MSSM
can be computed as functions of few parameters \cite{dnns};
Planck scale physics is employed only in the cancellation of the
cosmological constant and in the generation of the $R$-axion
mass \cite{axion}.
In this type of models, the lower limit on the $R$-axion mass
and the upper limit on the gravitino mass require the
supercolor scale to be in the range $10^2 - 10^4$ TeV \cite{dns}.

It is difficult to produce a large gluino mass in this type of models
because the supercolored fields cannot carry ordinary color.
In refs.~\cite{dn,dns,dnns} this problem is solved by an
indirect transmission of SUSY breaking, where the last link before the
MSSM is formed by vector-like quarks and leptons.

The mechanism for gluino mass generation presented in section 2
opens up the possibility of constructing simple visible sector models
along the lines of ref.~\cite{ads1}. In the rest of this
section we present some difficulties with this approach.

Consider a supercolor model with an unbroken global $U(1)$
symmetry which is gauged and identified with hypercharge.
As a result, all the scalars of the MSSM and the $G$-scalars
get mass at two loops proportional to their hypercharge \cite{ads1}.
Thus, the right-handed charged sleptons are the heaviest scalars
while the squark doublets are the lightest, the mass ratio
between these being six.
This might be a problem: depending on the
experimental lower bound for squarks (which is in the range $90 - 220$
GeV), the sleptons could be too heavy to insure the weak scale
stability (a recent study of the fine-tuning as a function of the
superpartner masses can be found in ref. \cite{nature}).

Since supercolored fields carry hypercharge, the photino
Majorana mass is easily produced.
The main problem encountered in ref.~\cite{ads1} is that
gluinos get only a very small mass at three-loop level from the
supercolor sector.
Here, however, a large gluino mass is produced by the $G$-color
sector supplemented by a gauge singlet superfield, as in section 3.
Note that in addition to the $S$ fermion exchange diagram,
there is a contribution to $G$-scalar mixing from a photino exchange.
The supercolored fields do not contribute to the mass
of the scalar singlet but this does not prevent $\tilde{S}$
from acquiring a VEV such that the electroweak breaking
occurs as in the hidden sector models.

The upper bound on the hypercharge of the $G$-colored superfields
(see eq~(\ref{eq:y})) implies that $M_{\tilde{\psi}_L}$ 
is at most half
of the right-handed charged slepton mass, which might be too low for
allowing a sufficiently large gluino mass.
If this is the case, $M_{\tilde{\psi}_L}$ 
can be enhanced by gauging a second
global symmetry of the supercolor sector and assigning its charges
to the $G$-scalars. Moreover, if the $G$-scalars become sufficiently
heavy, their two-loop contributions to the squark masses may reduce
the slepton to squark mass ratios.

The above considerations are based on the assumption that the two-loop
contributions of the supercolor sector give positive squared
squark and slepton masses.
The scalars (fermions) from the supercolor sector give negative
(positive) contributions \cite{dns,dn} so that at least some of the
fermions carrying hypercharge should be heavier than some
of the charged scalars.
This turns out to be a strong condition for supercolor models.
For example, the simplest supercolor sector, the $SU(3) \times SU(2)$
model \cite{ads1,axion}, is ruled out from this scheme since the
fermion component of the only light composite superfield charged under
$U(1)_Y$ is massless.

Also, the supercolor model should not allow a large Fayet-Iliopoulos
term for hypercharge at one loop since this would produce squark
VEV's \cite{dns,vev}. The $SU(6) \times U(1)$ model of
ref. \cite{dnns} is designed to cancel the Fayet-Iliopoulos term.
Again, in this model the charged fermions are lighter than the
scalars.

Yet another constraint is that all the charged fields should be
massive.
The $SU(5)$ model with two chiral superfields in the $\overline{5}$
representation and two in the 10 \cite{ads2,nou}, or the
$SU(3) \times SU(2)$ model  do not satisfy this condition.

However, it is quite possible that supercolor models with the
desired properties exist.
Only few of the known models of dynamical SUSY breaking were studied
in detail and, probably, new models will be find soon.
If a supercolor sector satisfying the above constraints is found,
a visible sector model including the $G$-color sector would
have certain advantages over the models of refs.~\cite{dnns,dns}:
SUSY breaking is transmitted
directly to the MSSM without need of additional fields
beyond the $G$-color sector and the singlet;
the supercolor scale is lower by an order of magnitude so that
the gravitino is lighter, satisfying easier the cosmological bounds
\cite{grav};
the role of the $\mu$-term is played successfully by the singlet.
Alternately, if the mechanisms for generating a $\mu$-term
discussed in refs. \cite{dnns,dns} remain problematic,
the models with indirect transmission of SUSY breaking may need
a $G$-color sector to give a weak scale VEV to a scalar singlet.

\section{Conclusions}

The complications in constructing realistic models of dynamical
SUSY breaking, mainly with producing a gluino mass and a $\mu$-term,
suggest a need for additional dynamics, beyond the one required to
break SUSY.

We have shown that a sufficiently large gluino mass can be
dynamically generated
if there are chiral superfields carrying both color and the charges
of an additional non-abelian gauge interaction, $G$-color.
When the $G$-colored scalars decouple, the $G$-color coupling constant
starts running faster and the $G$-fermions condense at a scale of
order 1 TeV. However, in order
to generate a gluino mass without fine-tuning, this link between
the $G$-color scale and the SUSY breaking scale should be very
effective, which happens only if the $G$-color $\beta$-function
nearly vanishes at scales above the scalar masses.
A nice feature of this mechanism is that the absence of the $\mu$-term
can be compensated naturally by the interactions of a gauge singlet.

Hidden sector models appear to become viable by including
this mechanism.
The only role of the supercolor sector in this case is to give
masses to the scalars in the visible sector and, therefore,
any model of dynamical SUSY breaking is adequate here.

By contrast,
simple visible sector models could be constructed using this mechanism
provided one finds a supercolor model which fulfills the following
conditions:
$i)$ some of the supercolored fields carry hypercharge;
$ii)$ the non-supersymmetric vacuum of the supercolor sector
preserves hypercharge;
$iii)$ in the effective low energy theory the charged fermions are
heavier than the charged scalars;
$iv)$ there is no large Fayet-Iliopoulos term for hypercharge;
$v)$ there is no massless charged field.

A more general issue which is relevant for this approach
is the behaviour of SUSY gauge theories  with
a scale of the order of the soft SUSY breaking terms.
This theories are in between the better understood
gauge theories with small soft SUSY breaking terms
and the non-SUSY gauge theories, and deserve more studies.

\section*{Acknowledgements}

I would like to thank Sekhar Chivukula, Lisa Randall, Martin Schmaltz
and John Terning for helpful discussions and useful suggestions.
{\em This work was supported in part by the National Science
  Foundation under grant PHY-9057173, and by the Department of Energy
  under grant DE-FG02-91ER40676.}


\vfil

\begin{thebibliography}{99}
\bibitem{ads1} I. Affleck, M. Dine and N. Seiberg, Nucl.\ Phys.\ {\bf
    B} {\bf 256}, 557 (1985).
\bibitem{dnns} M. Dine, A. E. Nelson, Y. Nir and Y. Shirman,
Report No. SCIPP 95/32, hep-ph/9507378.
\bibitem{ads2} I. Affleck, M. Dine and N. Seiberg, Phys.\ Rev.\ Lett.\
  {\bf 52}, 1677 (1984).
\bibitem{erich} E. Poppitz and S. P. Trivedi, report No. EFI-95-44
(1995), hep-th/9507169.
\bibitem{axion} J. Bagger, E. Poppitz and L. Randall, Nucl.\ Phys.\
  {\bf B} {\bf 426}, 3 (1994), hep-ph/9405345.
\bibitem{data} L. Montanet et al. (Particle Data Group), Phys.\ Rev.\
  {\bf D}
{\bf 50}, 1173 (1994) and 1995 partial update (http://pdg.lbl.gov/).
\bibitem{rsym} L. Hall and  L. Randall, Nucl.\ Phys.\ {\bf B}
 {\bf 352}, 289 (1991).
\bibitem{gluino} M. Dine and  D. A. MacIntire, Phys.\ Rev.\ {\bf D} 
{\bf 46}, 2594
(1992), hep-ph/9205227.
\bibitem{dsb} T. Banks, D. B. Kaplan and  A. E. Nelson,
Phys.\ Rev.\ {\bf D} {\bf 49}, 779 (1994), hep-ph/9308292.
\bibitem{dn} M. Dine and A. E. Nelson, Phys.\ Rev.\ {\bf D} 
{\bf 48}, 1277 (1993), hep-ph/9303230.
\bibitem{dns} M. Dine, A. E. Nelson and Y. Shirman, Phys.\ Rev.\ {\bf
    D} {\bf 51}, 1362 (1995), hep-ph/9408384.
\bibitem{tc} S.~Dimopoulos and L.~Susskind, Nucl.\ Phys.\ {\bf B} 
{\bf 155}, 237 (1979);\\
E.~Eichten and K.~Lane, Phys.\ Lett.\ {\bf B} {\bf 90}, 125 (1980);\\
E.~H.~Simmons, Nucl.\ Phys.\ {\bf B} {\bf 312}, 253 (1989).
\bibitem{oneloop} R. Barbieri, L. Girardello and A. Masiero,
Phys.\ Lett.\ {\bf B} {\bf 127}, 429 (1983); \\
R. Barbieri and L. Maiani, Nucl.\ Phys.\ {\bf B} 
{\bf 243}, 429 (1984).
\bibitem{ads3} I. Affleck, M. Dine and N. Seiberg, Nucl.\ Phys.\ {\bf
    B} {\bf 241}, 493 (1984).
\bibitem{exotic} O. Aharony, M. E. Peskin, J. Sonnenschein and
S. Yankielowicz, preprint SLAC-PUB-6938 July 1995, hep-th/9507013.
\bibitem{link} A.~Kagan and S.~Samuel, Phys.\ Lett.\ {\bf B} 
{\bf 252} (1990) 605.
\bibitem{susytc}  S.~Samuel, Nucl.\ Phys.\ {\bf B} 
{\bf 347}, 625 (1990); \\
A.~Kagan, in {\it Proceedings of the 15th Johns Hopkins
    Workshop on Current Problems in Particle Theory}, ed. by
  G.~Domokos and
  S.~Kovesi-Domokos (World Scientific, Singapore, 1992), p.217; \\
B.~A.~Dobrescu, Nucl.\ Phys.\ {\bf B} 
{\bf 449}, 462 (1995), hep-ph/9504399.
\bibitem{nat} G.~'t Hooft, in {\it Recent Developments in Gauge
Theories}, ed. by G.~'t Hooft, {\it et al.} (Plenum, New York, 1980),
p.135.
\bibitem{twoloop} D. R. T. Jones, Phys.\ Rev.\ {\bf D} 
{\bf 25}, 581 (1982).
\bibitem{walk} B. Holdom, Phys.\ Rev.\ {\bf D} 
{\bf 24}, 1441 (1981); Phys.\ Lett.\ {\bf B} {\bf 150}, 301 (1985); \\
K. Yamawaki, M. Bando and K. Matumoto, Phys.\ Rev.\ Lett.\ 
{\bf 56}, 1335 (1986);\\
T. Appelquist, D. Karabali and L.C.R. Wijewardhana, Phys.\ Rev.\
Lett.\ {\bf 57}, 957 (1986);\\
T. Appelquist and L.C.R. Wijewardhana, Phys.\ Rev.\ {\bf D} 
{\bf 35}, 774 (1987);
Phys.\ Rev.\ {\bf D} {\bf 36}, 568 (1987).
\bibitem{ewsb} L. Ibanez and G. Ross, Phys.\ Lett.\ {\bf B} 
{\bf 110}, 215 (1982);\\
L. Alvarez-Gaume, M. Claudson and M. B. Wise, Nucl.\ Phys.\ {\bf B} 
{\bf 207}, 96
(1982).
\bibitem{higgs} See e.g.,
J. F. Gunion and H. E. Haber, Nucl.\ Phys.\ {\bf B} 
{\bf 272}, 1 (1986).
\bibitem{singlet} J. Bagger, E. Poppitz and L. Randall,
report No. EFI-95-21 (1995), hep-ph/9505244.
\bibitem{wall} For a review, see: A. Vilenkin, Phys.\ Rep.\ 
{\bf 121}, 263 (1985).
\bibitem{decay} J. Preskill, S. P. Trivedi, F. Wilczek and
M. B. Wise, Nucl.\ Phys.\ {\bf B} {\bf 363}, 207 (1991).
\bibitem{nature} G. W. Anderson and D.J. Castano,
Phys.\ Rev.\ {\bf D} {\bf 52}, 1693 (1995), hep-ph/9412322.
\bibitem{vev} L. Alvarez-Gaume, J. Polchinski and M. B. Wise,
Nucl.\ Phys.\ {\bf B} {\bf 221}, 495 (1983).
\bibitem{nou} T. A. ter Veldhuis, report No. VAND-TH-95-7 (1995),
hep-th/9510121.
\bibitem{grav} H. Pagels and J. R. Primack, Phys.\ Rev.\ Lett.\ 
{\bf 48}, 223 (1982).

\end{thebibliography}
\end{document}